\documentclass[%
 aip,
 amsmath,amssymb,
 reprint,%
]{revtex4-1}

\usepackage{graphicx}
\usepackage{dcolumn}
\usepackage{bm}

\usepackage[utf8]{inputenc}
\usepackage[T1]{fontenc}
\usepackage{mathptmx}
\usepackage{etoolbox}

\usepackage{booktabs}
\usepackage{multirow}

\makeatletter
\def\@email#1#2{%
 \endgroup
 \patchcmd{\titleblock@produce}
  {\frontmatter@RRAPformat}
  {\frontmatter@RRAPformat{\produce@RRAP{*#1\href{mailto:#2}{#2}}}\frontmatter@RRAPformat}
  {}{}
}%
\makeatother
\begin{document}

\preprint{AIP/123-QED}

\title{Low-barrier-height Hf$_\text{1-x}$Zr$_\text{x}$O$_2$ ferroelectric tunnel junctions with large ON-state currents}

\author{Shaoqing Ding}
\author{Jinyuan Yao}
\affiliation{\mbox{Department of Physics, The Pennsylvania State University, University Park, Pennsylvania 16802, USA}}
\affiliation{\mbox{Materials Research Institute, The Pennsylvania State University, University Park, Pennsylvania 16802, USA}}

\author{Eknath Sarkar}
\author{Emmanuel Quezada}
\affiliation{\mbox{School of Electrical and Computer Engineering, Georgia Institute of Technology, Atlanta, Georgia 30332, USA}}

\author{Thomas N. Jackson}
\affiliation{\mbox{School of Electrical Engineering and Computer Science, The Pennsylvania State University, University Park,}\\\mbox{ Pennsylvania 16802, USA}}
\affiliation{\mbox{Materials Research Institute, The Pennsylvania State University, University Park, Pennsylvania 16802, USA}}

\author{Susan Trolier-McKinstry}
\affiliation{\mbox{Department of Materials Science and Engineering, The Pennsylvania State University, University Park,}\\\mbox{Pennsylvania 16802, USA}}
\affiliation{\mbox{Materials Research Institute, The Pennsylvania State University, University Park, Pennsylvania 16802, USA}}

\author{Suman Datta}
\affiliation{\mbox{School of Electrical and Computer Engineering, Georgia Institute of Technology, Atlanta, Georgia 30332, USA}}

\author{Ying Liu*}
\affiliation{\mbox{Department of Physics, The Pennsylvania State University, University Park, Pennsylvania 16802, USA}}
\affiliation{\mbox{Materials Research Institute, The Pennsylvania State University, University Park, Pennsylvania 16802, USA}}
\email{yxl15@psu.edu}

\date{\today}

\begin{abstract}
Ferroelectric tunnel junctions (FTJs) hold great promise for nonvolatile memory, neuromorphic computing, and other applications. Hf$_\text{1-x}$Zr$_\text{x}$O$_2$ (HZO) has become a leading candidate for the tunnel barrier due primarily to its compatibility with CMOS technology. Recent advances on HZO based FTJs have highlighted the importance of understanding the electrical transport through a ferroelectric tunnel barrier for fulfilling the technological promises of FTJs. In this paper we report investigation of HZO FTJs featuring a barrier of ultrathin HZO ($\leq$5nm) measured down to a temperature of 2 K. Direct quantum tunneling was found to dominate at low temperatures, along with contributions from trap-assisted tunneling. As the temperature increases, conduction by thermal activation takes over. We developed a minimal model that accounts for all experimental observations, revealing a barrier height significantly lower than that of the band alignment picture, an unexpected finding attributable to the emergence of a shallow trap band. This finding reveals a previously unexplored pathway toward simultaneously achieving large ON-state currents, high tunnel electroresistance and improved temperature stability for the high-speed memory applications of FTJs.
\end{abstract}

\maketitle

The explosive growth in AI and other data-rich applications of microelectronics call for next-generation devices and architectures that will help break the bottleneck of high-speed computing and the energy consumption associated with moving data between storage and logic functions \cite{theis_end_2017}. Ferroelectric tunnel junctions (FTJs) possessing a ferroelectric tunnel barrier sandwiched between a top and bottom electrode (TE and BE) provide a promising solution to this grand challenge\cite{Esaki_Polar_1971,kohlstedt_theoretical_2005,zhuravlev_giant_2005}. FTJs, which feature two distinct current states corresponding to the two directions of the ferroelectric polarization in the barrier to form the ON and OFF states, can function as nonvolatile memory elements that supports high-density, low-power data storage with nondestructive readout \cite{garcia_ferroelectric_2014}. FTJs  are expected to also facilitate compute-in-memory (CiM) \cite{sebastian_memory_2020} and neuromorphic computing  \cite{chanthbouala_ferroelectric_2012,boyn_learning_2017}.

The performance parameters of FTJs must be optimized for these applications. First, those desired for all modern microelectronic memories, including robust retention time ($>10$ years), endurance ($>10^{10}$ cycles), temperature stability, and CMOS compatibility\cite{das_beyond_2021,yoon_review_2023} must be achieved. In addition, the tunnel electroresistance (TER) ratio defined as $(J_\text{ON} - J_\text{OFF})$$/$$J_\text{OFF}$, where $J_\text{ON}$ and $J_\text{OFF}$ are current densities of the ON and OFF states and $J_{\text{ON}}$ itself must be as large as possible\cite{chu_ultra-thin_2021}.

To optimize the performance parameters, the mechanism controlling the electrical transport through a ferroelectric barrier must be understood. Several models have been used previously to gain insights into the process \cite{velev_predictive_2016}. Quantum tunneling (QT) based models include the direct and Fowler-Nordheim (F-N) QT \cite{garcia_giant_2009,gruverman_tunneling_2009,velev_predictive_2016,park_ferroelectric_2024,fowler_electron_1928} as well as indirect or trap-assisted tunneling (TAT) \cite{svensson_trapassisted_1973,makram-ebeid_quantum_1982,cheng_nitridationenhanced_1988}. Thermal activation (TA) based models include Poole-Frenkel (P-F) emission in which trapped electrons are thermally excited into the conduction band \cite{poole_dielectric_1916,frenkel_pre-breakdown_1938} and thermionic emission (TE) of electrons from a metal \cite{richardson_some_1912,dushman_electron_1923}. (See Supplementary Materials (SM) \cite{SM} for more discussion on these models). It was rare, however, that the current density ($J$) $vs$. the electric field ($E$) characteristics of a FTJ were analyzed in the entirety with the values of the parameters obtained from the analysis examined in detail\cite{park_ferroelectric_2024}.

In this work we focus on the electrical transport through an ultrathin ferroelectric barrier of CMOS compatible Hf$_{0.5}$Zr$_{0.5}$O$_2$ (HZO) with a thickness of $\leq$5nm to allow direct QT. To simplify the problem, symmetrical metallic electrodes were used without optimizing the TER ratio. The FTJs are measured from a temperature ($T$) as low as 2 K to one as high as 400\,K to allow $J$ through the FTJs be dominated by QT at low- and TA at high-$T$ limits. A minimal model capable of describing all features seen experimentally was developed. The surprising discovery we made in this work is that the tunnel barrier height is lowered significantly from that expected from the band alignment picture, which we attribute to the formation of a defect band, revealing a pathway towards high-performance FTJs not considered previously.

\section{Experimental}

The preparation of the FTJ structures started with a heavily p-type doped ($\rho$<0.005 $\Omega$$\cdot$cm) Si wafer. To remove the native oxide from the surface, buffer oxide etchant (BOE) was used to allow for good backside contact. The bottom tungsten (W) electrode (20\,nm thick) was deposited via physical vapor deposition (PVD). Plasma enhanced atomic layer deposition (PEALD) was used to deposit the 3.5 and 5\,nm thick ferroelectric Hf$_{0.5}$Zr$_{0.5}$O$_2$ (HZO) films. The precursors used for HfO$_2$ and ZrO$_2$ were tetrakis(dimethylamino)hafnium (TDMAH) and tetrakis(dimethylamino)zirconium (TDMAZ), respectively. The thin films were deposited at 250$^{\circ}$C and process pressure of 900\,mTorr with O$_2$ plasma power of 600\,W. To ensure a 1:1 Hf $vs$. Zr ratio, alternating layers of HfO$_2$ to ZrO$_2$ were deposited with argon as the source gas for intermediate purge steps. An \textit{in situ} ellipsometer was used to monitor the film thickness during deposition.

The bilayer films of W/HZO were capped with 80\,nm W via sputtering and then patterned with a maskless aligner. To define the top electrode with lateral dimensions ranging from 50 to 150\,$\mu$m, the W was etched with reactive ion etching in SF$_6$ plasma at 150\,W. After etching, the photoresist was stripped in a stripper solution at 85$^{\circ}$C. Finally, rapid thermal annealing (RTA) was used to treat the HZO film. The temperatures and times treating the 5\,nm and 3.5\,nm films are 550$^{\circ}$C for 1 min and 650$^{\circ}$C for 30 seconds, respectively. Both films were annealed in N$_2$.

The devices were further processed for low-temperature measurements by fabricating expanded top contacts. An insulating layer of 100\,nm Al$_2$O$_3$ was evaporated and patterned by photolithography via lift-off to isolate the substrate while maintaining access to the top electrodes. Finally, metal stacks of 5\,nm Ti and 100\,nm Au were deposited, which were defined by lift-off to ensure electrical connection during low temperature measurement.

Room-temperature polarization $vs$. electric field ($P$-$E$) hysteresis and Positive-Up-Negative-Down (PUND) polarization measurements were performed in a FormFactor 11000 probe station using W probes to establish electrical contact with the top and bottom electrodes. Low-temperature $P$-$E$ and PUND measurements were performed in a LakeShore CRX-VF cryogenic probe station. Data acquisition was carried out using a Keithley 4200-SCS Semiconductor Characterization System equipped with a Pulse Measure Unit (4225-PMU) to generate the required waveforms. The $P$-$E$ hysteresis loops were measured using a triangular waveform at a frequency of 25\,kHz. Before any measurement, the devices were woken up by continuous cycling using the same waveform as the $P$-$E$ hysteresis measurements with a maximum amplitude of 3\,MV/cm until the polarization reached its maximum value.

For current density $vs$. electric field ($J$-$E$) measurements, the samples were mounted onto a puck of the Physical Property Measurement System (PPMS, Quantum Design). Electrical connections to the devices were established by pressing In dots onto the electrodes and connecting them to the puck's contact pads using gold wires. The puck was then loaded into for the low temperature measurements. DC transport measurements were conducted using a Keithley 2400 SourceMeter to source voltage and measure the resulting current from 2 to 400 K.

\section{HZO as a ferroelectric tunnel barrier}

Films of HZO is a leading FTJ barrier material due to its compatibility with CMOS processing. ALD or PEALD grown and rapid thermal annealing (RTA) treated ultrathin films of HZO can be made ferroelectric at the nanometer thicknesses needed for QT \cite{boscke_ferroelectricity_2011,cheema_enhanced_2020,cheema_one_2022}. 
For relatively thick HZO barriers ($\geq 7$\,nm), electrical transport was found to be dominated by P-F emission\cite{park_study_2015,mallick_oxygen_2023} in which the Coulomb potential barrier for the trapped electrons is lowered by the application of field to increase the current through the barrier. Given that P-F emission does not involve QT, its suitability for describing the transport data raises the question as to whether the devices with a thick barrier should even be called tunneling ones. In contrast, QT was found to play an important role in FTJs featuring an ultrathin barriers as seen Table\,S1 in the SM \cite{SM}). So far, the main fucus of work in this area has been to achieve a large TER through the optimizing barrier profile, the use of asymmetric electrodes, including the combination of metallic and semiconducting electrodes, and the engineering of interfacial layers \cite{park_ferroelectric_2024}.

Quantitative modeling of the $J$-$E$ characteristics was performed previously within the framework of QT, reproducing the main features seen in the data \cite{gruverman_tunneling_2009,ambriz-vargas_complementary_2017,yoon_understanding_2019}. However, only room-temperature $J$-$E$ data for individual HZO thicknesses were analyzed, typically by using many adjustable parameters, including, for example, an overall prefactor, up to two interfacial and tunnel barrier heights, the optical dielectric constant, and the effective mass of the tunneling electron. The dependence of the current on the thickness and the temperature, as well as the validity of the obtained fitting parameters were not examined in detail. The leakage currents associated with defects in the barrier were typically not treated either.


\begin{figure}[htbp!]
\includegraphics[width=0.46\textwidth]{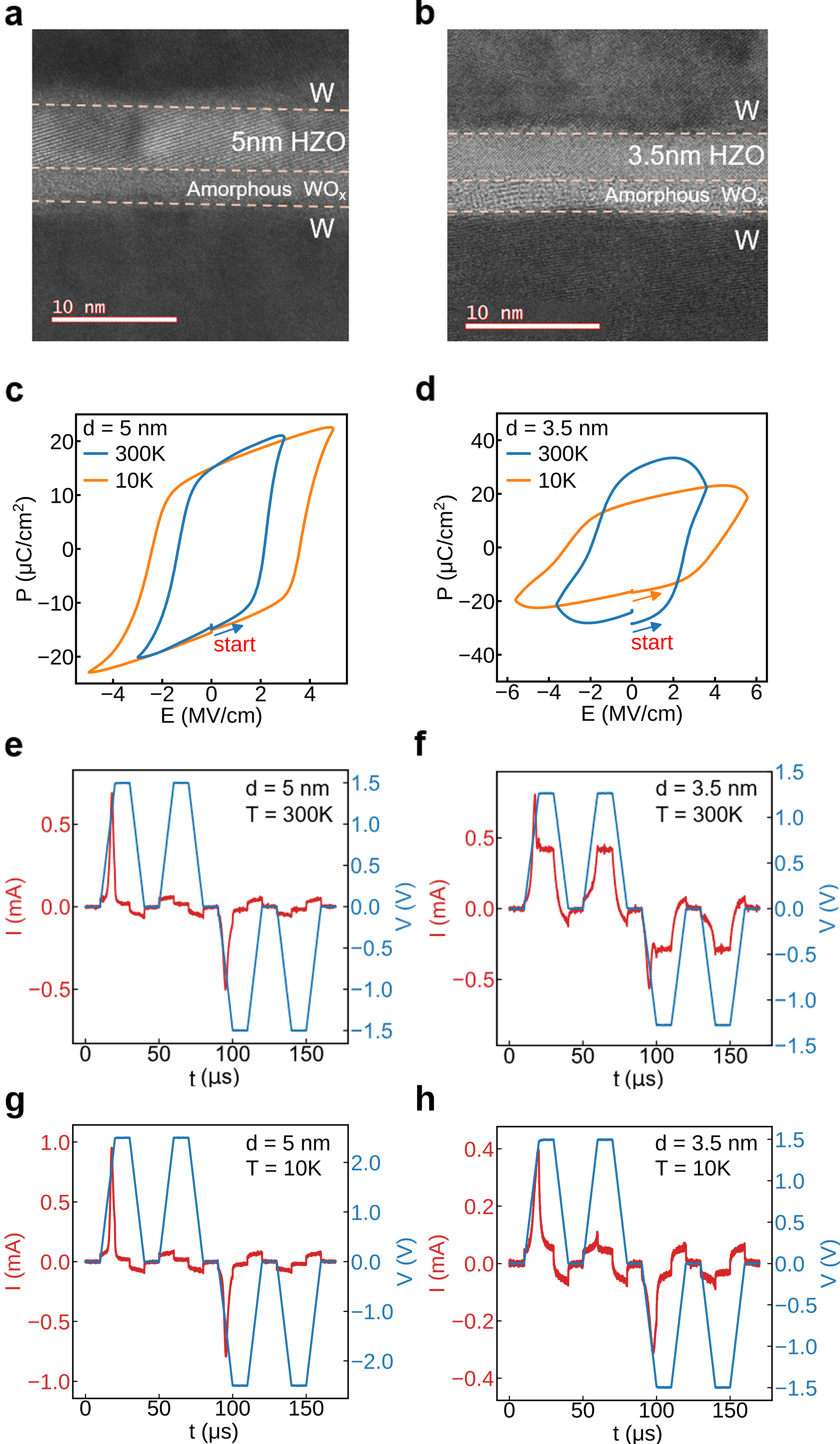}
\caption{\label{fig:1} \textbf{Structural and electrical characterization of W/HZO/W FTJs.}
\textbf{a}, \textbf{b}, Cross-sectional TEM images of W/HZO/W stacks with, respectively, a 5\,nm (\textbf{a}) and 3.5\,nm (\textbf{b}) thick Hf$_{0.5}$Zr$_{0.5}$O$_2$ (HZO). An amorphous WO$_x$ interfacial layer is visible between the HZO and the BE for both HZO thicknesses. Bars indicate the length scales. The thicknesses of the WO$_x$ layers are 2.6\,nm for the 5\,nm and 2.4\,nm for the 3.5\,nm HZO devices, respectively.
\textbf{c}, \textbf{d}, polarization $vs$. electric field ($P$–$E$) hysteresis loops for the 5\,nm (\textbf{c}) and 3.5\,nm (\textbf{d}) HZO devices measured at room temperature (300 K, blue) and 10 K (orange). Both barriers were seen to exhibit robust ferroelectricity even though the room-temperature loop for the 3.5\,nm barrier is inflated by the presence of the leakage current.
\textbf{e}-\textbf{h}, Positive-Up-Negative-Down (PUND) measurements at room temperature (\textbf{e}, \textbf{f}) and 10 K (\textbf{g}, \textbf{h}) for the 5\,nm (\textbf{e}, \textbf{g}) and 3.5\,nm (\textbf{f}, \textbf{h}) HZO devices. The applied voltage sequences (blue) and resulting current responses (red) clearly differentiate the ferroelectric switching peaks from the leakage currents.
}
\end{figure}

In the current work, the FTJs consist of W BE and TE and a PEALD grown tunnel barrier of 5 or 3.5\,nm thick HZO treated by RTA as done previously\cite{Mittmann_impact_2020,materano_influence_2020} Figs.\,\ref{fig:1}a and b show cross-sectional transmission electron microscopy (TEM) images of the W/HZO/W stacks in both 5 and 3.5\,nm devices. The HZO thicknesses are consistent with the number of PEALD growth cycles. An amorphous WO$_x$ interfacial layer between HZO and the BE was found, which was expected due to the use of oxygen plasma for the HZO growth. The  We also performed room- and low-temperature $P$-$E$ and PUND measurements on the HZO. These measurements verified good ferroelectric properties of the HZO, yielding a remanent polarization of approximately 15\,$\mu$C\,cm$^{-2}$ for the 5\,nm barrier at the room temperature and 14.9\,$\mu$C\,cm$^{-2}$ at 10 K (Fig.\,\ref{fig:1}c). The 3.5\,nm barrier is seen to show an inflated $P$-$E$ loop and failure to close the loop at room temperature, both of which are signs of increased leakage current, as expected. This inflation was suppressed at 10 K due to the suppression of the leakage current, yielding a remanent polarization of approximately 16.8\,$\mu$C\,cm$^{-2}$ (Fig.\,\ref{fig:1}f). To estimate the coercive fields for barriers of both thicknesses, the voltage drops across the WO$_x$ interfacial layer, which were found to be 2.6\,nm for the 5\,nm and 2.4\,nm for the 3.5\,nm HZO, respectively, out of total bias voltages of 1.5 and 2.5 for the 5\,nm and 1.3 and 1.5 V for the 3.5\,nm HZO devices at room and low temperatures, respectively (Figs.\,\ref{fig:1}c-f). If we treat the WO$_x$ layer as conductive as W, which will likely to be true at room temperature but likely off at 10 K due to the electronic localization in an amorphous film, the coercive fields are estimated to be about 2\,MV\,cm$^{-1}$ at room temperature and up to 3.5\,MV\,cm$^{-1}$ at 10 K (ignoring the voltage drop across the WO$_x$ layer) for the 5\,nm HZO (Fig.\,\ref{fig:1}c). For 3.5\,nm HZO, the coercive field was difficult to estimate at room temperature because of the inflated $P$-$E$ loop and the uncertainty of the voltage drop across the WO$_x$ layer at 10 K. However, it is clear that the coercive field increases as the temperature is lowered.   

\section{Results on transport through ultrathin HZO}

\begin{figure}[htbp!]
\includegraphics[width=0.45\textwidth]{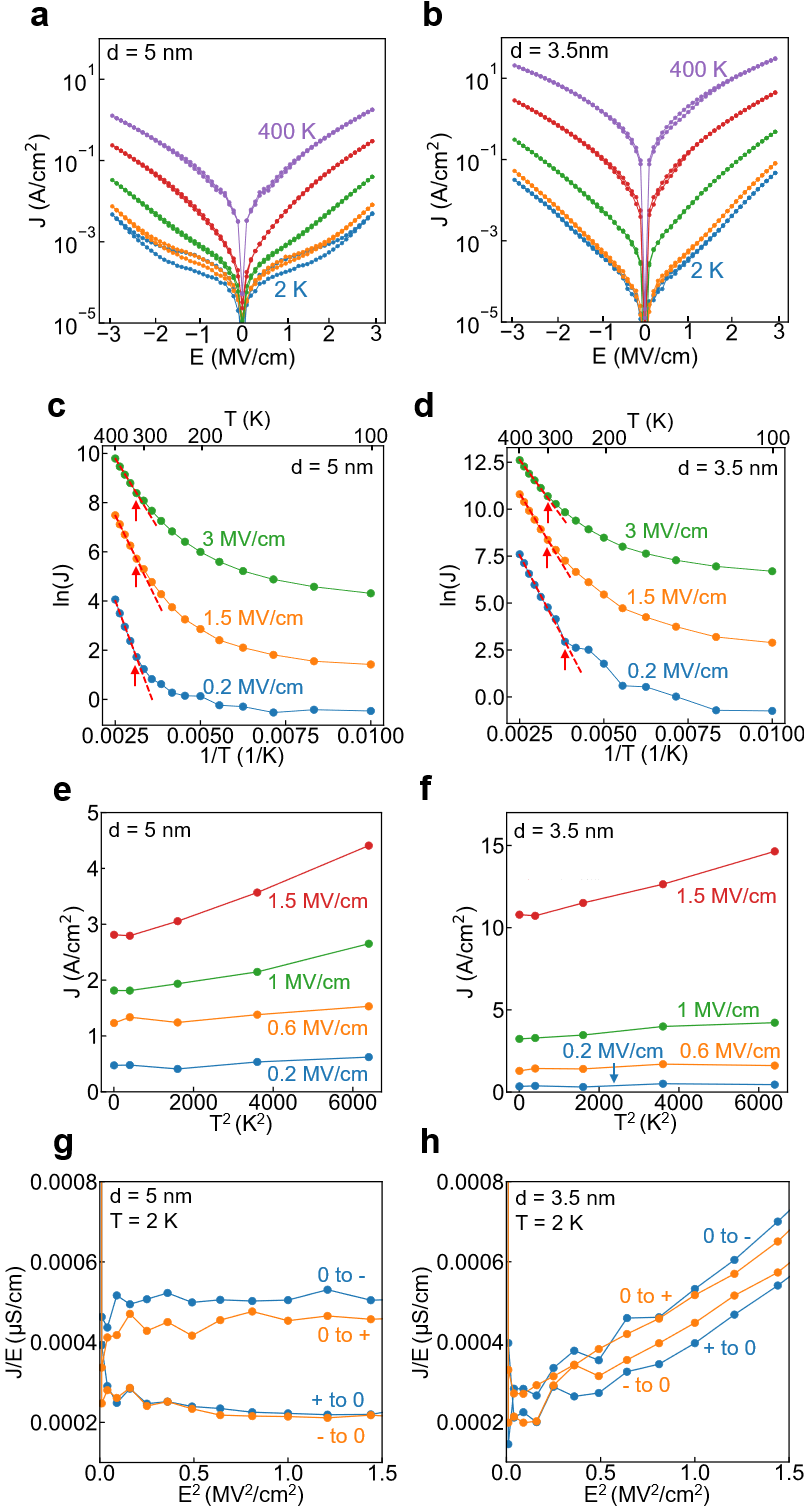}
\caption{\label{fig:2} \textbf{Electrical transport measurements.}
\textbf{a}, \textbf{b}, Current density ($J$) $vs$. electric field ($E$) curves measured at a temperature ($T$) of 2, 100, 200, 300, and 400\,K for a 5 (\textbf{a}) and 3.5\,nm (\textbf{b}) thick HZO device, respectively. \textbf{c}, \textbf{d}, Arrhenius plots of $\ln J$ $vs$. $1/T$ for the 5\,nm (\textbf{c}) and 3.5\,nm (\textbf{d}) HZO devices for $E$ = 0.2, 1.5, and 3\,MV/cm, with the TA energies obtained from the Arrhenius fit 0.32, 0.24, and 0.19\,eV for the 5\,nm and 0.30, 0.25, and 0.20\,eV for the 3.5\,nm HZO devices, respectively. Arrhenius behavior was seen above 320\,K for the 5\,nm and 300\,K for the 3.5\,nm HZO device. 
\textbf{e}, \textbf{f}, $J$ $vs$. $T^2$ at selected electric fields (0.2, 0.6, and 1) for the 5\,nm (\textbf{e}) and 3.5\,nm (\textbf{f}) HZO devices from $T$ = 2 to 80\,K. The linear behavior seen in the low-$T$ limit is consistent with the theoretical expectations.
\textbf{g}, \textbf{h}, Conductivity ($J/E$) $vs$. $E^2$ for the 5 (\textbf{g}) and 3.5\,nm (\textbf{h}) devices at 2\,K.
}
\end{figure}

We measured the $J$-$E$ characteristics of W/HZO/W FTJs with a 5\,nm or 3.5\,nm HZO thick barrier from 400\,K down to 2\,K (Figs.\,\ref{fig:2}a and b) with the maximal bias voltages being 1.5  for the former and 1.05 V for the latter barrier. The values of the electric field are calculated with the voltage drops across the WO$_x$ layer ignored. Several trends are easily identified. First, values of $J$ at the same temperature ($T$) and $E$ obtained from the 3.5\,nm HZO devices are found to be significantly higher than those of the 5\,nm ones in the high-temperature limit. At low temperatures, however, the difference in $J$ values between the two thicknesses at the same $T$ and $E$ is less significant.  Second, below 100\,K, $J$ is seen to increase exponentially with $E$ in the large $E$ limit. Furthermore, the onset of $E$ over which the trend is seen is smaller in devices with a thinner HZO - approximately 0.5\,MV/cm for the 3.5\,nm and $\sim$2\,MV/cm for the 5\,nm HZO device. Finally, $J$ at fixed $T$ varied little below 100\,K but increases rapidly when $T$ $\geq$ 200\,K. 

These trends, seen consistently in multiple devices (see SM for additional data), suggest strongly that the electrical transport through the ultrathin HZO is dominated by QT in the low-$T$ limit. The $\ln J$ $vs$. $E$ behavior is consistent with QT picture in which the tunnel barrier is modulated by the application of $E$. Only when $T$ becomes sufficiently high does the TA take over. In the high-$T$ limit, the Arrhenius behavior, $\ln J$ vs. $1/T$ for fixed $E$, was observed as shown in Figs.\,\ref{fig:2}c and d, confirming that the TA processes dominate the electrical transport. Importantly, the barrier height over which the thermally activated electron must overcome can be extracted from fitting the Arrhenius behavior to the data in the high-$T$ limit at small field, which turned out to be around 0.31\,eV for all devices. It should be noted, however, that the Arrhenius plots only provide a rough estimate of the barrier potential height (see below). Finally, the overall trend of the difference of currents as a function of $T$ and $E$ for the two thicknesses of the barrier indicate that whether the conduction through barrier is interface or bulk dominated is less than straightforward (see SM for more discussion).

Interestingly, even $J$ from pure QT can still be $T$-dependent due to the Fermi-Dirac distribution of the electrons in the electrodes, which leads to the $J$ $vs$. $T^2$ behavior \cite{stratton_volt-current_1962}. This expectation is consistent with experiment as shown in Figs.\,\ref{fig:2}e and f. According to the theory, $J(T\neq0)/J(0)=1+\frac{1}{6}(\pi c k_B T)^2$, where $c$ is a constant that depends on the barrier profile. In Figs.\,\ref{fig:2}g and h, $J/E$ is plotted against $E^2$ in the low-$E$ regime. The ohmic term alone appears to capture the behavior of the 5\,nm HZO device, whereas a cubic term is clearly present in the 3.5\,nm HZO device. However, the exponential dependence quickly takes over as $E$ increases. The crossover from direct QT to F-N tunneling is not seen, suggesting that the tunnel barrier in these devices does not simply evolve from a trapezoidal to triangular profile within the experimental range of $E$ values (see SM for more discussion). Absence of crossover was also seen in previous studies of FTJs\cite{hwang_physics_2024}. 

\begin{figure}[htbp!]
\includegraphics[width=0.45\textwidth]{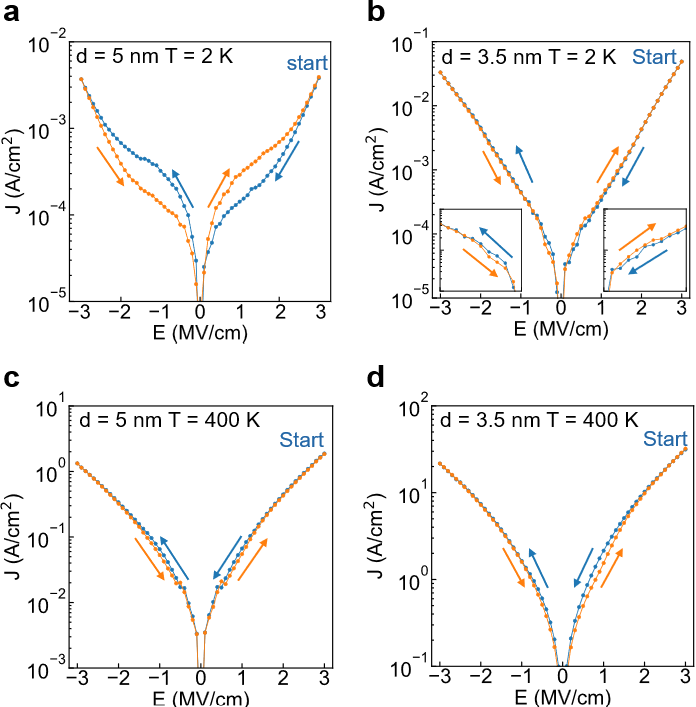}
\caption{\label{fig:3} \textbf{Temperature-dependent hysteresis loops.}
\textbf{a}, \textbf{b}, Current density $vs$. electric field ($J$-$E$) hysteresis loops measured at 2\,K for devices having a 5\,nm (\textbf{a}) and 3.5\,nm (\textbf{b}) thick barrier of HZO. The low-current state is obtained when sweeping the field from $\pm$3\,MV/cm to 0. The high-$J$ states were seen when $E$ was swept from 0 to $\pm$3\,MV/cm. The 3.5\,nm device exhibits the same direction of the hysteresis loop even though the difference in $J$ is small (see insets in \textbf{b} for zoomed-in views of the hysteresis).
\textbf{c}, \textbf{d}, $J$-$E$ hysteresis loops measured at 400\,K for the 5\,nm (\textbf{c}) and 3.5\,nm (\textbf{d}) HZO devices. The direction of the hysteresis loops is seen to reverse for both thicknesses such that the high-$J$ states were obtained in the sweep from $+$3\,MV/cm to $-$3\,MV/cm while the low-$J$ state from $-$3\,MV/cm to $+$3\,MV/cm. The measurement of each hysteresis loop was completed with 112 seconds. 
}
\end{figure}

Hysteresis is seen in the $T$-dependent $J$-$E$ curves. In the low-$T$ limit, the low-$J$ branch was obtained when the $E$ was swept from a large magnitude to zero, whereas the high-$J$ branch was obtained in the opposite sweep (Figs.\,\ref{fig:3}a-b), opposite to that commonly seen in FTJs. Interestingly, at high-$T$ limit, the direction in the hysteresis loop, which is tiny but present, is seen to reverse (Figs.\,\ref{fig:3}c-d), recovering the behavior found in FTJs with a relatively thick HZO. The interesting question is whether the reversal of the hysteresis loop direction as the temperature is lowered is due to the absence of the ferroelectric switching or the low rate at which the bias voltage was swept, which tends to make the ferroelectric switch invisible in the hysteresis loop. To clarify this point, current $vs$. bias voltage curves were measured with different sweep rates at three different temperatures (400, 300, and 213 K) for devices of both the 5\,nm and 3.5\,nm thick HZO. As seen in Figs. S10-15 in SM, ferroelectric switching was clearly seen in the 5\,nm HZO device around the bias voltage of 0.5 V for sweep rates $\geq$ 100 V/s at 213 K. The sweep rate used for $J$-$E$ measurements in Fig. \ref{fig:2}a were lower than the lowest rates in Figs. S10-15, suggesting that the ferroelectric switching was present at high temperatures. However, as seen Figs.\,\ref{fig:1}c-d, the ferroelectric switching is unlikely to have taken place with the range of the bias voltage (below the coercive field) with which the hysteresis loops were obtained at the lowest temperatures. Thus, the ferroelectric switching is unrelated to the hysteresis in the 5\,nm HZO at low temperatures. The same conclusion holds for devices possessing a 3.5\,nm thick HZO.


Given that the hysteresis in 5\,nm HZO device is much more prominent than the one in 3.5\,nm HZO device, the physical origin of the hysteresis must be related to traps whose effect is most prominent in barrier barrier ($e.g.$, the P-F conduction). Since the conduction in the low-temperature limit is dominant by quantum tunneling, trap assisted tunneling (TAT), which is an indirect quantum tunneling process in which an electron first tunnels into a trap within the barrier from one electrode, then tunnels out from that trap to another electrode, provides an addition channel for the electrical transport through the barrier of HZO and a natural explanation for the unusual direction of the hysteresis loop in the $J$-$E$ loops (see below). 


\section{Minimal model for the electrical transport}

The experimental results presented above suggest that at the minimum, three separate conduction channels through ultrathin HZO are present, contributing a current density of $J_{\mathrm{QT}}$ from QT, $J_{\mathrm{TA}}$ from TA, and $J_{\mathrm{TAT}}$ from trap-assisted conduction, the TAT (see below), with the total current density through HZO $J_{\mathrm{tot}}$, given by
\begin{equation}
J_{\mathrm{tot}}(\mathcal{E},T) = J_{\mathrm{TA}}(\mathcal{E},T) + J_{\mathrm{QT}}(\mathcal{E},T) + J_{\mathrm{TAT}}(\mathcal{E},T),
\label{eq:Jsum_lowT}
\end{equation}
This will be a minimal model for electrical transport through an ultrathin HZO. 

To proceed, we start with a barrier potential denoted by \(U(\mathcal{E},x)\). The profile of the potential will evolve as the the bias voltage increases. The ferroelectric polarization of the tunnel barrier will modify the barrier potential through electrostatics quantified by $\Delta U_{\text{pol}}(x)$. In the present study, however, the strong screening in metallic electrodes will make the effect of ferroelectric polarization vanishingly small. As pointed out above, heavily doped WO$_3$ can be treated as metallic, as reported previously\cite{Altendorf_metallicWO3_2016, Leng_metallicWO3_2017}. This assumption is consistent with the small TER ratios seen in the $J$-$E$ curves (see SM).

The effect of image charge represented by $\Delta U_{\text{im}}(x)$, where the x axis is perpendicular to the junction plane with the origin at the interface between the BE and HZO, is considered. Putting all these together, we have

\begin{equation}
U(\mathcal{E},x)=U_0+\Delta U_{\text{im}}(x)-q\mathcal{E}x.
\label{eq:Barrier}
\end{equation}

The height of the potential barrier is defined by the maximum of \(U(\mathcal{E},x)\). 

In principle, $\Delta U_{\text{im}}(x)$ should reflect an infinite series of image charges due to the presence of two parallel metal surfaces. We will however truncate this series to include only the first-order positive and negative image charges as the contributions from higher-order terms decay rapidly due to the screening from the mobile charge inside the HZO (see below for more discussion), making the effective screening length, $\lambda_D$, essentially the Debye length, very small \cite{onsager_surface_1934,simmons_generalized_1963}. We then have


\begin{eqnarray}
\Delta U_{\text{im}}(x) =&& -\frac{e^2}{16\pi \epsilon_r \epsilon_0} \biggl[ \frac{\exp[-2x/\lambda_D]}{x} \nonumber \\&& + \frac{\exp[-2(d-x)/\lambda_D]}{d-x} \biggr],
\label{eq:Image_Charge}
\end{eqnarray}

where $\epsilon_r$ is the optical dielectric constant and $d$ the barrier thickness with $\lambda_D$ given by \cite{debye_theory_1923}
\begin{equation}
\lambda_D = \sqrt{\epsilon_{s}\epsilon_0 k_B T / e^2 n}.
\label{eq:Debye length}
\end{equation}

$\epsilon_{s}$, which accounts for the redistribution of mobile charge of density $n$, should be the static dielectric constant.

The current density from QT, $J_{\mathrm{QT}}$, is given by
\cite{bardeen_tunnelling_1961,harrison_tunneling_1961,stratton_volt-current_1962}:

\begin{eqnarray}
J_{\mathrm{QT}}(\mathcal{E},T) =&& J_{0,\mathrm{QT}} \frac{e m_{\mathrm{1}}}{2\pi^{2}\hbar^{3}} \nonumber\\ &&\times \int_{0}^{E_{\max}} t(\mathcal{E},E_x,T) N(\mathcal{E},E_x,T)dE_x.
\label{eq:JQT_landauer}
\end{eqnarray}

where $J_{0,\mathrm{QT}}$ is a unitless normalization constant, $e$ the charge of the electron, $m_{\mathrm{1}}$ the effective electron mass in the electrodes, $t(\mathcal{E},E_x,T)$ the tunneling probability, $N(\mathcal{E},E_x,T)$ the supply function, and $E_x$ the longitudinal kinetic energy in the conduction band of the electrode, with the integration from $E_x=0$ to the peak of the potential barrier, $E_{\max}$. Using the WKB approximation \cite{wentzel_verallgemeinerung_1926,kramers_wellenmechanik_1926,Brillouin_mecanique_1926}, $t$ for the effective potential $U(\mathcal{E},x)$ is given by

\begin{equation}
t(\mathcal{E},E_x) = \exp \left[ -\frac{2}{\hbar} \int_{0}^{d} \sqrt{2m_{\mathrm{2}}[U(\mathcal{E},x) - E_x]} dx \right],
\label{eq:Ttunnel_WKB}
\end{equation}
where $m_{\mathrm{2}}$ is the effective mass of the electron in HZO. 

The supply function $N(\mathcal{E},T,E_x)$ incorporates the Fermi-Dirac distribution functions of BE and TE under $E$. Integrating the transverse kinetic energies ($E_y$, $E_z$), we have

\begin{equation}
N(\mathcal{E},E_x,T) = k_BT\ln \left[ \frac{1+\exp \left( -\frac{E_x-\mu}{k_B T} \right)} {1+\exp \left( -\frac{E_x-\mu+e\mathcal{E}d}{k_B T} \right)} \right],
\label{eq:supply_function}
\end{equation}
with $\mu$ the chemical potential of the BE ($x \leq 0$).

For conduction by TA, although P-F emission provides a reasonable fit to the high-temperature $J$–$E$ curves, the fitting requires that the value of the dielectric constant, $\epsilon_r$, be much larger than that for HZO (see SM), which suggests that P-F emission is not the appropriate mechanism for electrical transport through the ultrathin HZO even at high temperatures. On the other hand, an electron gaining sufficient amount of thermal energy to overcome a potential barrier, which does not involve a trap inside HZO, is very similar to the emission of electrons from a metal. Therefore, it is reasonable to adopt the Richardson-Dushman (R-D) equation \cite{richardson_some_1912,dushman_electron_1923} for $J_{\mathrm{TA}}(\mathcal{E},T)$ to describe the TA contribution. The R-D equation reads 

\begin{equation}
J_{\mathrm{TA}}(\mathcal{E},T) = J_{0,\mathrm{TA}} \frac{q m_{\mathrm{1}} k_B^2}{2\pi^2 \hbar^3} T^2 \exp \left( -\frac{U_{\mathrm{TA}}(\mathcal{E})}{k_B T} \right).
\label{eq:JTA_thermionic}
\end{equation}
where $J_{0,\mathrm{TA}}$ is a unitless fitting prefactor. The activation energy $U_{\mathrm{TA}}(\mathcal{E})$ corresponds to the maximum height of $U(\mathcal{E},x)$ defined in Eq.\,(\ref{eq:Barrier}). Crucially, because $U(\mathcal{E},x)$ incorporates the image charge potential, $U_{\mathrm{TA}}(\mathcal{E})$ will be $E$-dependent.

Contribution from the TAT process, denoted by
\(J_{\mathrm{TAT}}\), can be calculated from the formula for indirect tunneling of electrons from BE to TE through a localized trap state \cite{svensson_trapassisted_1973,makram-ebeid_quantum_1982,cheng_nitridationenhanced_1988}, which is of the form (see SM)
\begin{eqnarray}
J_{\mathrm{TAT}}(\mathcal{E},T)
=&&
J_{0,\mathrm{TAT}}\int_{0}^{E_\text{max}}\int_{0}^{d}
\frac{t_L t_R}{t_L+t_R} \nonumber\\
&&\times N(\mathcal{E},E_x,T)  n_t(x,E_x)\,dx\,dE_x,
\label{eq:TAT}
\end{eqnarray}
where $t_L$ and $t_R$ are the tunneling matrix elements of the two tunneling steps in TAT through a trap located at $x$, and $n_t(x,E_x)$ denotes the energy and positional distribution of the traps. It should be noted that the TAT current is determined by the properties of the traps, which are not affected very much by ferroelectricity of the barrier. 

Since the sweeping rate of the bias voltage used in this experiment is low, we expecte no charge accumulates in the barrier. As a result, the product of the tunneling matrix elements, $t_L$ and $t_R$ should depend only on the thickness of HZO if the width of trap site itself is neglected. Since $t_Lt_R$ is a constant, $t_Lt_R/(t_L+t_R)$ is maximized when $t_L=t_R$, corresponding to equal barrier thicknesses for two tunneling steps, which is achieved when trap site is in the center of the barrier, as discussed previously\cite{cheng_nitridationenhanced_1988}.

The above picture explains the clockwise hysteresis loop observed at low temperatures. When a high bias voltage is applied, the barrier potential is strongly tilted. The deep traps close to the interface where the barrier potential is mostly lowered have a high probability to be occupied, while the traps near the opposite side remain mostly unoccupied. As the available traps concentrate off the center of the barrier, the rate of TAT reduces. Thus, a low-$J$ branch is achieved when $E$ is ramped down from a high value. When sweep from low toward high $E$, trap depopulation takes place, more traps available near the center of the barrier become available for TAT, leading to the high-current branch of the hysteresis loop.

Finally, both direct QT and TAT in the low-$E$ limit may be expanded into the odd power terms of $E$ (see SM). For convenience in numerical modeling, the TAT contribution is approximated by 
\begin{equation}
J_{\mathrm{3}}(\mathcal{E},T) = \sigma_{\text{3}}(T) \mathcal{E} + \gamma_{\text{3}}(T) \mathcal{E}^3,
\label{eq:Jtrap_ohmic}
\end{equation}
where $\sigma_{\text{3}}(T)$ and $\gamma_{\text{3}}(T)$ are treated as fitting parameters.

\section{Precision test of the minimal model}

\begin{figure}[htbp!]
\includegraphics[width=0.45\textwidth]{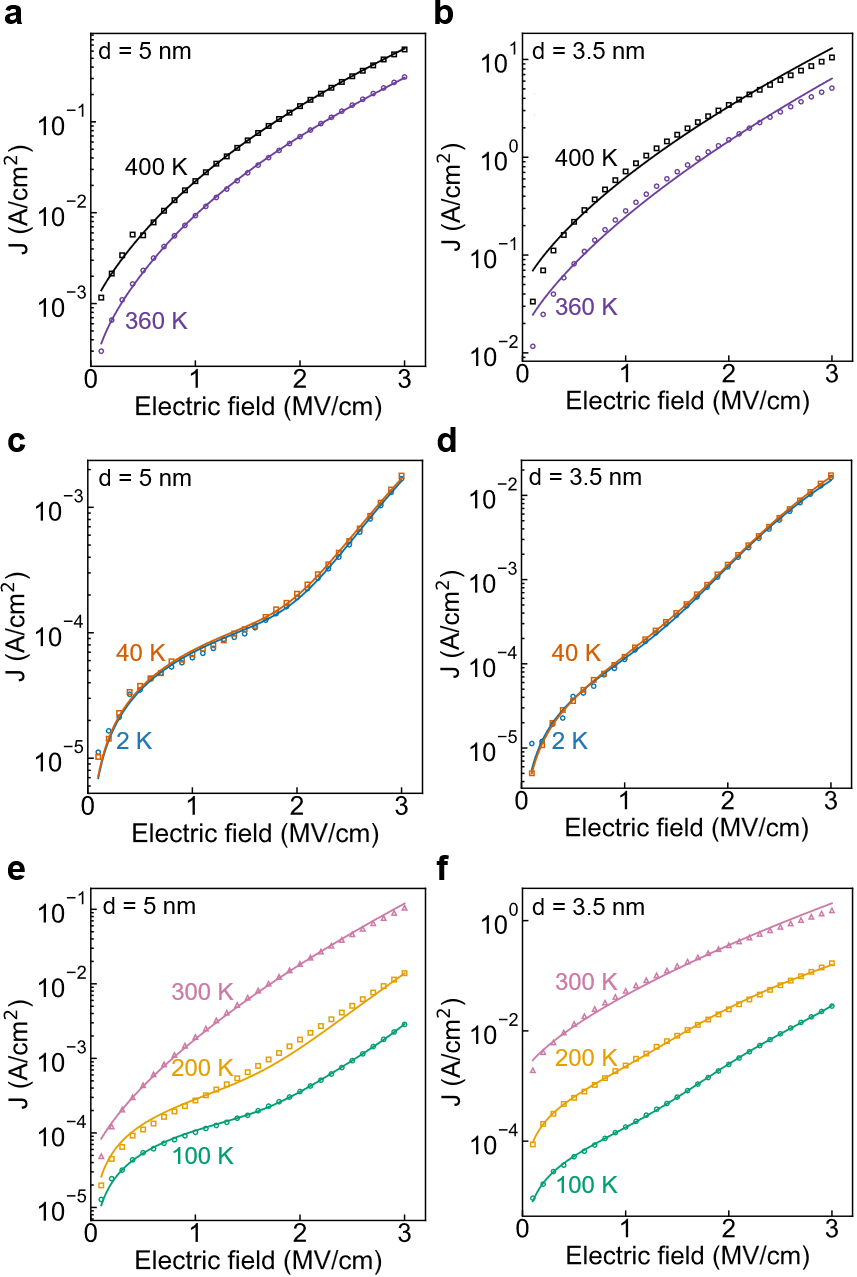}
\caption{\label{fig:4} \textbf{Fitting the three-channel model to the $J$-$E$ data.}
\textbf{a}, \textbf{b}, $J$-$E$ curves and corresponding fits (solid lines) for the 5\,nm (\textbf{a}) and 3.5\,nm (\textbf{b}) HZO devices at 360\,K and 400\,K including contributions from TA and TAT.
\textbf{c}, \textbf{d}, $J$-$E$ curves and corresponding model fits for the 5\,nm (\textbf{c}) and 3.5\,nm (\textbf{d}) HZO devices at 2\,K and 40\,K including contributions from QT and TAT.
\textbf{e}, \textbf{f}, $J$-$E$ curves for the 5\,nm (\textbf{e}) and 3.5\,nm (\textbf{f}) HZO devices measured at intermediate temperatures (100\,K, 200\,K, and 300\,K) using the full three-channel model.
}
\end{figure}

To test the minimal model, we take advantage of a wide range of temperature over which the data was collected. In the high-$T$ limit, $J$ is dominated by the TA contribution (Figs.\,\ref{fig:2}a and b), which is controlled mainly by the potential height of the barrier akin to the interface-limited electrical transport (see SM for more discussion). Numerical modeling suggests that fitting to the data improves if contribution from TAT is also included. It is remarkable that only five and four parameters are needed in the modeling of the 3.5 and 5\,nm HZO devices, respectively, to obtain high-quality fitting (Figs.\,\ref{fig:4}a and b). The small difference between the fit and the data in the high-$E$ limit of the 3.5\,nm device is likely due to a non-negligible contribution from direct QT, which may also explain the low TA energies obtained at high $E$ in the Arrhenius plots, which yielded TA energies lower than the $U_{0,\mathrm{TA}}$ values obtained from the modeling, 0.43 eV for the 3.5\,nm and 0.48 eV 5\,nm thick HZO device.

In the low-$T$ limit, the electrical transport through ultrathin HZO is dominated by QT and TAT, making it possible to reduce the number of fitting parameters by ignoring the TA contribution. Tunneling process is strongly influenced by the barrier thickness, which explains, qualitatively, the bulk-limited features we found from the difference between $J$ at the same $E$ of 5\,nm and 3.5\,nm HZO devices at low temperatures (see SM). Based on data shown in Fig.\,\ref{fig:2}g, Eq.\,\ref{eq:Jtrap_ohmic} will be truncated to keep only the Ohmic term for the 5\,nm HZO samples with the $\gamma_{\text{3}}(T)$ term kept only for the 3.5\,nm HZO device (Fig.\,\ref{fig:2}h). 
As a result, the number of fitting parameters is four for the 5\,nm device and five for the 3.5\,nm one. It should be noted that $U_{0,\mathrm{QT}}$ and $m_2/m_0$ affect the potential profile in a correlated way. We therefore performed extensive modeling searching for fitting parameters so that the value of $m_2/m_0$ obtained from the multi-parameter fitting is consistent with those found in the literature \cite{yoon_understanding_2019,fontanini_charge_2022}, which further improves the reliability of the value of $U_{0,\mathrm{QT}}$ obtained from the modeling.

Fitting Eqs.\,(5) and (10) to the $J$-$E$ data obtained from 2 to 40 K results in strong agreement between the minimal model and the data as shown in Figs.\,\ref{fig:4}c and d. Values of the fitting parameters obtained in the low- and high-$T$ limits are summarized in Table\,\ref{tab:fitting_params_final}. To fit the data obtained at intermediate temperatures, contributions from all three channels are included, yielding high-quality fitting shown in Figs.\,\ref{fig:4}e and f, as expected. The fitting parameters obtained in the intermediate temperatures are found in the SM. We note that the small normalization factors obtained from the modeling comes from the simple need to match the units in theoretical formulas and those used experimentally and the sample related properties. What matters is the relative size of the contributions from two different channels, not the absolute values of the prefactors. The small prefactors resulted from the numerical modeling do not affect the determination of the values of the barrier height and the effective mass.

\begin{table}[htbp!]
\centering
\caption{\label{tab:fitting_params_final}
\textbf{Values of fitting parameters obtained from the modeling.}
Parameters used to fit the minimal model to the experimental data include the normalization factors $J_{0,\text{QT}}$ and $J_{0,\text{TA}}$ which are used to account for the units, the sample size, and other sample related factors, $U_{0,\text{QT}}$ and $U_{0,\text{TA}}$ are the barrier heights used in the QT and TA equations, $m_2/m_0$ is ratio of the effective mass of electrons in HZO to that in vacuum, $\lambda_D$ is the screening length in HZO, and $\sigma_{\text{3}}$ and $\gamma_{\text{3}}$ are two parameters characterizing conduction through TAT (see Eq.\,(10)).}

\renewcommand{\arraystretch}{1.4}
\begin{tabular}{llcc}
\toprule
\textbf{Temperature} &
\textbf{Parameter} &
\textbf{5\,nm device} &
\textbf{3.5\,nm device} \\
\midrule

\multirow{6}{*}{\textbf{2\,K}}
& $J_{0,\mathrm{QT}}$             & $7.7\times10^{-9}$  & $1.3\times10^{-8}$ \\
& $U_{0,\mathrm{QT}}$ (eV)        & 0.61                 & 0.50 \\
& $m_2/m_0$                       & 0.25                 & 0.25 \\
& $\sigma_3$                      & $6.9\times10^{-11}$ & $4.9\times10^{-11}$ \\
& $\gamma_3$                      & --                   & $1.3\times10^{-24}$ \\

\midrule

\multirow{6}{*}{\textbf{400\,K}}
& $J_{0,\mathrm{TA}}$             & $3.5\times10^{-5}$  & $3.9\times10^{-4}$ \\
& $U_{0,\mathrm{TA}}$ (eV)        & 0.48                 & 0.43 \\
& $\lambda_D$ (nm)                & 1.4                  & 0.85 \\
& $\sigma_3$                      & $3.5\times10^{-16}$ & $1.2\times10^{-11}$ \\
& $\gamma_3$                      & --                   & $1.0\times10^{-28}$ \\

\bottomrule
\end{tabular}
\end{table}

\section{Lowering of the barrier height}

\begin{figure}[htbp!]
\includegraphics[width=0.45\textwidth]{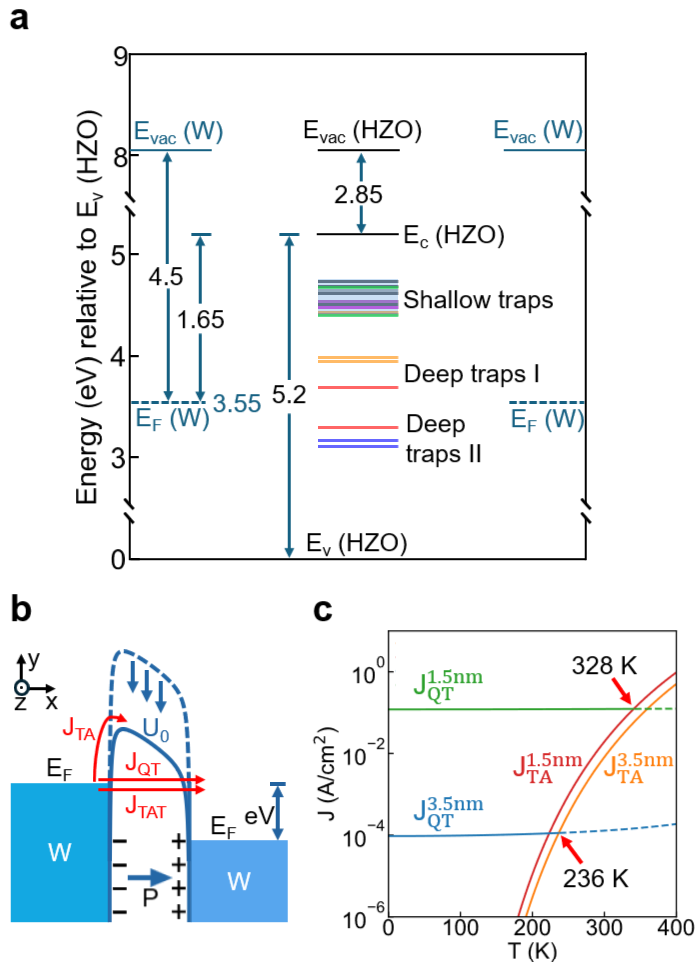}
\centering
\caption{\label{fig:5} \textbf{Origin and consequences of barrier height lowering.}
\textbf{a}, Energy band diagram of W/HZO/W at zero bias showing the landscape of oxygen vacancy levels relative to the valence band maximum (VBM), $E_\text{v}$, which is the reference point for all energies shown. This choice of reference point differs from that used in the transport model. Shallow deep traps of Groups I and II are shown relative the the Fermi energy of W. The vacuum levels of W and HZO are aligned. The conduction band minimum (CBM) for HZO, $E_\text{c}$, is also shown. 
\textbf{b}, Schematic of the lowering of the barrier height along with the three conduction channels of the minimal model (see text). Ferroelectric polarization, P, of the tunnel barrier, P, is also illustrated.
\textbf{c}, Predicted $J$-$T$ curves with the currents from QT and TA at 1\,MV/cm of 3.5\,nm and 1.5\,nm HZO. The curves are computed using the values of $U_{0,\text{TA}}$ and $U_{0,\text{QT}}$ obtained from the modeling (Table\,\ref{tab:fitting_params_final}). The crossover temperatures for the lowered barrier height for 3.5\,nm and 1.5\,nm HZO devices are labeled.
}
\end{figure}

The values of $U_{0,\text{TA}}$ and $U_{0,\text{QT}}$ obtained from the modeling, shown in Table\,\ref{tab:fitting_params_final}, are between 0.42 - 0.61 eV. To begin with, the barrier potential, \(U(\mathcal{E},x)\), is defined by the energy of the tunneling electron in the barrier at which the kinetic energy approaches zero. In the band alignment picture, the vacuum levels of $all$ materials are aligned. The barrier potential is obtained by taking the Fermi energy of BE to be zero, that of TE offset by the bias voltage, and the conduction band minimum (CBM) forms \(U(\mathcal{E},x)\) since the kinetic energy of the tunneling electron is nonzero only above CBM. As shown in Fig.\,\ref{fig:5}a, $U_{0}$ should be 1.65 eV, much higher than those obtained from the modeling, which suggests that the barrier height is lowered from the band alignment picture despite of the variations in the $U_{0,\text{TA}}$ and $U_{0,\text{QT}}$ values.

The unexpected finding of low barrier height suggests that the tunneling electrons in HZO does not need to be in the conduction band to have nonzero kinetic energy (be mobile). An additional band must have formed below the CBM in which the electrons are mobile at a temperature as low as 2 K. Recall that HZO used in the present study was treated by RTA. It is natural then to link the proposed additional band with that originating from oxygen vacancies. Interestingly, a defect band below CBM, referred to as a "shallow trap band (STB)", was already demonstrated in 20\,nm thick HZO grown by ALD through photoemission electron microscopy (PEEM) and photoluminescence spectroscopy (PL) studies\cite{vega_modifications_2025}. It was shown that laser dose can be used to modulate the local concentration of oxygen vacancies \cite{beechem_photoinduced_2024}. The STB of a width 0.5 eV was found to roughly 0.8 eV below CBM. The Fermi level in the HZO, which is determined by the filling of the STB, was found to decrease with the decreasing vacancy concentration controlled by the laser dose. Another study on oxygen vacancies in ALD-grown HZO thin films also suggested a dominant defect level around 0.6 eV below the CBMfrom P-F fitting of $J$-$E$ curves and PL spectra of HZO\cite{mallick_oxygen_2023}.

In addition, first-principle calculations together with experimental measurements\cite{perevalov_electronic_2018,wei_spontaneous_2019,chai_first-principles_2022, mallick_oxygen_2023, yang_mathrmdftuv_2025, vega_modifications_2025, noor_oxygen_2026} revealed a large number of states associated with oxygen vacancies group in a small energy range roughly 1 eV below the CBM of HZO (Fig.\,\ref{fig:5}a). Other trap states at lower energies, which we refer to as shallow and Group I and II deep traps, were also found. Even though the energies of defect levels in ultrathin HZO were not determined directly in the present work and therefore may deviate from those reported in literature, these results summarized in Fig.\,\ref{fig:5}a support strongly the formation of the STB in HZO so long that the concentration of the oxygen cacancies is sufficiently high. 

In this regard, the concentration of the oxygen vacancies in ultrathin HZO was shown previously to be as high as $7\times10^{19}\,\text{cm}^{-3}$\, \cite{engl_peak_2025}, exceeding significantly the Mott criterion\cite{mott_conduction_1969}, $N_c \approx 6.8 \times 10^{18}\,\text{cm}^{-3}$ for HZO, the threshold at which the overlap of the wave functions of the defect states form a "defect band", along with a mobility edge\cite{Neamen_SM_2012}. Based on these previous experimental results, the lowest energy of the defect states will be 0.6 eV above the Fermi energies of W, the value of $U_{0,\text{QT}}$ for the 5\,nm device and within 0.1 eV of that of the 3.5\,nm device obtained from the modeling (Table\,\ref{tab:fitting_params_final}). Values of $U_{0,\text{TA}}$ for both the 3.5 and 5\,nm devices, slightly lower than those of $U_{0,\text{QT}}$, are within 0.2 eV of that expected in this picture, which may be due to lowering of the mobility edge \cite{Neamen_SM_2012}.

\section{Discussion and Conclusions}

The results presented above provide insights into improving the performance of FTJs. Reducing the barrier height in FTJs is analogous to using semiconducting Si as the barrier in Josephson junctions, which are key components in quantum technologies. The barrier height was lowered from roughly 2.9~eV in Nb/Al$_2$O$_3$/Nb to 0.3~eV in Nb/Si/Nb Josephson junctions, providing options for superconducting circuit design\cite{xiong_silicon-based_2026}. In the current work, the main advantage brought by the lowering of the barrier height is an increased ON-state current density. Although the ON-state current density of FTJs suitable for microelectronics applications is not a universal value but one depending on circuitry design such as array size, interconnect resistance, sensing architecture, and selector characteristics\cite{luo_correlated_2024}, a larger On-state current is strongly desirable. To investigate the effect of a lowered barrier height, we simulated the $J_{QT}$ with 1.65 and 0.85~eV $U_0$ using the fitting parameters obtained from our 3.5\,nm HZO device data (Fig.~S9 in SM). We found that the predicted ON-state $J$ increases by three orders of magnitude from lowering the barrier height, which can reduce greatly the read-out delay and enable faster FTJ operations. The effect of reducing the HZO thickness from 3.5 to 1.5~nm (Fig.~\ref{fig:5}c) can also be calculated similarly, leading to a predicted ON-state possessing also a three orders of magnitude increase. In addition, QT is predicted to be dominant well above room temperature (328~K) in 1.5\,nm HZO device, which is favorable for temperature stable operation.

However, lowering $U_0$ will also increase the OFF-state $J$. Similarly, TAT, an indirect quantum tunneling process which was found to enable the Josephson diode effect\cite{ding_indirect_2026}, is expected to contribute to the OFF-state $J$, resulting in unwanted energy consumption. Integrating a nonlinear selector with the FTJ can suppress the OFF-state current as demonstrated using a selector in an ultrathin HZO FTJ\cite{jung_superior_2023}, with the selector performance co-optimized further with suitable memory and array designs\cite{luo_correlated_2024}. The OFF-state current can also be reduced while maintaining a high ON-state $J$ by a large TER ratio as well. It has been shown that asymmetric electrodes are effective in increasing TER\cite{garcia_giant_2009,gruverman_tunneling_2009,velev_predictive_2016,park_ferroelectric_2024}, an approach demonstrated explicitly in the current work (see SM). Here, polarization-dependent depletion or accumulation of charge carriers modulates the barrier profile, resulting in an enhanced TER. Alternatively, a large TER may be obtained by using an appropriate metal electrode to form a ferroelectric Schottky diode\cite{yao_ferroelectric_2023}.

Combining barrier-height reduction with engineering of the interface and electrode provides a pathway toward simultaneously achieving high read-out current, large TER, and enhanced temperature stability. In this context, oxygen vacancies are not purely defects with widely recognized detrimental consequences, such as degraded retention, endurance, and device uniformity, but can also serve as a resource for enhancing FTJ performance. Nevertheless, several challenges must be addressed towards this goal, including field-induced vacancy migration and the tendency of vacancies to redistribute, which may cause imprint, temporal drift, and conductive-filament formation. Oxygen vacancies therefore need to be carefully managed using tools made available in recent years, ranging from first-principles searches for vacancy states to ALD growth and RTA design strategies, as well as modulation of the concentration and spatial distribution of oxygen vacancies by laser processing\cite{beechem_photoinduced_2024,vega_modifications_2025}.

In conclusion, we fabricated W/HZO/W FTJs featuring a ferroelectric barrier of a 3.5 or 5\,nm thick HZO and measured them from 2\,K to 400\,K. A minimal model that accounts for all experimental observations in a quantitative manner was developed and tested numerically, which revealed unexpectedly a barrier height lowering from the band alignment picture. We suggest that the lowering is due to the emergence of the STB. These findings delineate a new pathway to pursue high-performance FTJs through oxygen vacancy engineering not considered previously.

\section*{Supplementary Material}
Additional data and technical information are included in the Supplementary Material.

\begin{acknowledgments}
This work is supported by the US Department of Energy, Office of Science, Office of Basic Energy Sciences Energy Frontier Research Centers program under Award Number DE-SC0021118.
\end{acknowledgments}

\section*{Author Declarations}

\subsection*{Conflict of Interest}
The authors have no conflicts to disclose.

\subsection*{Author Contributions}
\textbf{Shaoqing Ding}: Data curation (lead); Formal analysis (lead); Conceptualization (supporting); Writing – original draft (lead).
\textbf{Jinyuan Yao}: Conceptualization (supporting); Writing – review \& editing (supporting).
\textbf{Eknath Sarkar}: Data curation (supporting); Formal analysis (supporting).
\textbf{Emmanuel Quezada}: Data curation (supporting); Formal analysis (supporting).
\textbf{Thomas N. Jackson}: Conceptualization (supporting).
\textbf{Susan Trolier-McKinstry}: Conceptualization (supporting); Writing – review \& editing (supporting).
\textbf{Suman Datta}: Conceptualization (supporting); Methodology (supporting).
\textbf{Ying Liu}: Conceptualization (lead); Methodology (lead); Formal analysis (lead); Writing – original draft (lead).

\section*{Data Availability Statement}
The data that support the findings of this study are openly available in \url{https://app.globus.org/file-manager/collections/41189b1f-8fe1-47f6-96c1-fb40e4299a0f/sharing}

\bibliography{HZO2}

@misc{SM,
note = {See Supplemental Material for additional electrical characterization results and theory background.}
}

@article{richardson_some_1912,
	title = {Some applications of the electron theory of matter},
	volume = {23},
	journal = {Philos. Mag.},
	author = {Richardson, O.W.},
	year = {1912},
	pages = {594--627},
}

@article{poole_dielectric_1916,
	title = {On the dielectric constant and electrical conductivity of mica in intense fields},
	volume = {32},
	journal = {Philos. Mag. },
	author = {Poole, H.H.},
	year = {1916},
	pages = {112--129},
}

@article{debye_theory_1923,
  author  = {Debye, Peter and H{\"u}ckel, Erich},
  title   = {Zur Theorie der Elektrolyte. I. Gefrierpunktserniedrigung und verwandte Erscheinungen},
  journal = {Phys. Z.},
  volume  = {24},
  pages   = {185--206},
  year    = {1923}
}

@article{dushman_electron_1923,
	title = {Electron emission from metals as a function of temperature},
	volume = {21},
	journal = {Phys. Rev.},
	author = {Dushman, Saul},
	year = {1923},
	pages = {623--636},
}

@article{wentzel_verallgemeinerung_1926,
	title = {Eine Verallgemeinerung der Quantenbedingungen für die Zwecke der Wellenmechanik},
	volume = {38},
	journal = {Z. Phys.},
	author = {Wentzel, Gregor},
	year = {1926},
	pages = {518--529},
}

@article{kramers_wellenmechanik_1926,
	title = {Wellenmechanik und halbzahlige Quantisierung},
	volume = {39},
	journal = {Z. Phys.},
	author = {Kramers, H. A.},
	year = {1926},
	pages = {828--840},
}

@article{Brillouin_mecanique_1926,
	title = {La mécanique ondulatoire de {Schrödinger}: une méthode générale de resolution par approximations successives},
	volume = {183},
	journal = {C. R. Acad. Sci.},
	author = {Brillouin, Léon},
	year = {1926},
	pages = {24--26},
}

@article{fowler_electron_1928,
	title = {Electron emission in intense electric fields},
	volume = {119},
	journal = {Proc. R. Soc. Lond. A},
	author = {Fowler, Ralph Howard and Nordheim, L.},
	year = {1928},
	pages = {173--181},
}

@article{onsager_surface_1934,
	title = {The surface tension of {Debye}--{Hückel} electrolytes},
	volume = {2},
	journal = {J. Chem. Phys.},
	author = {Onsager, Lars and Samaras, Nicholas N. T.},
	year = {1934},
	pages = {528--536},
}

@article{frenkel_pre-breakdown_1938,
	title = {On pre-breakdown phenomena in insulators and electronic semi-conductors},
	volume = {54},
	journal = {Phys. Rev.},
	author = {Frenkel, J.},
	year = {1938},
	pages = {647--648},
}

@article{bardeen_tunnelling_1961,
	title = {Tunnelling from a many-particle point of view},
	volume = {6},
	journal = {Phys. Rev. Lett.},
	author = {Bardeen, J.},
	year = {1961},
	pages = {57--59},
}

@article{harrison_tunneling_1961,
	title = {Tunneling from an independent-particle point of view},
	volume = {123},
	journal = {Phys. Rev.},
	author = {Harrison, Walter A.},
	year = {1961},
	pages = {85--89},
}

@article{stratton_volt-current_1962,
	title = {Volt-current characteristics for tunneling through insulating films},
	volume = {23},
	journal = {J. Phys. Chem. Solids},
	author = {Stratton, R.},
	year = {1962},
	pages = {1177--1190},
}

@article{simmons_generalized_1963,
	title = {Generalized formula for the electric tunnel effect between similar electrodes separated by a thin insulating film},
	volume = {34},
	journal = {J. Appl. Phys.},
	author = {Simmons, John G.},
	year = {1963},
	pages = {1793--1803},
}

@article{mott_conduction_1969,
	title = {Conduction in non-crystalline materials: {III}. Localized states in a pseudogap and near extremities of conduction and valence bands},
	volume = {19},
	journal = {Philos. Mag.},
	author = {Mott, N. F.},
	year = {1969},
	pages = {835--852},
}

@article{Esaki_Polar_1971,
	title = {Polar switch},
	volume = {13},
	journal = {IBM Tech. Discl. Bull.},
	author = {Esaki, L. and Laibowitz, R.B. and Stiles, P.J.},
	year = {1971},
	pages = {2161--2162},
}

@article{svensson_trapassisted_1973,
	title = {Trap-assisted charge injection in {MNOS} structures},
	volume = {44},
	journal = {J. Appl. Phys.},
	author = {Svensson, Christer and Lundström, Ingemar},
	year = {1973},
	pages = {4657--4663},
}

@article{makram-ebeid_quantum_1982,
	title = {Quantum model for phonon-assisted tunnel ionization of deep levels in a semiconductor},
	volume = {25},
	journal = {Phys. Rev. B},
	author = {Makram-Ebeid, S. and Lannoo, M.},
	year = {1982},
	pages = {6406--6424},
}

@article{cheng_nitridationenhanced_1988,
	title = {Nitridation-enhanced conductivity behavior and current transport mechanism in thin thermally nitrided {$\mathrm{SiO}_{2}$}},
	volume = {63},
	journal = {J. Appl. Phys.},
	author = {Cheng, X. R. and Cheng, Y. C. and Liu, B. Y.},
	year = {1988},
	pages = {797--802},
}

@article{zhuravlev_giant_2005,
	title = {Giant electroresistance in ferroelectric tunnel junctions},
	volume = {94},
	journal = {Phys. Rev. Lett.},
	author = {Zhuravlev, M. Ye. and Sabirianov, R. F. and Jaswal, S. S. and Tsymbal, E. Y.},        
	year = {2005},
	pages = {246802},             
}

@article{kohlstedt_theoretical_2005,
	title = {Theoretical current-voltage characteristics of ferroelectric tunnel junctions},
	volume = {72},
	journal = {Phys. Rev. B},
	author = {Kohlstedt, H. and Pertsev, N. A. and Rodríguez Contreras, J. and Waser, R.},
	year = {2005},
	pages = {125341},
}

@article{garcia_giant_2009,
	title = {Giant tunnel electroresistance for non-destructive readout of ferroelectric states},
	volume = {460},
	journal = {Nature},
	author = {Garcia, V. and Fusil, S. and Bouzehouane, K. and Enouz-Vedrenne, S. and Mathur, N. D. and Barthélémy, A. and Bibes, M.},
	year = {2009},
	pages = {81--84},
}

@article{gruverman_tunneling_2009,
	title = {Tunneling electroresistance effect in ferroelectric tunnel junctions at the nanoscale},
	volume = {9},
	journal = {Nano Lett.},
	author = {Gruverman, A. and Wu, D. and Lu, H. and Wang, Y. and Jang, H. W. and Folkman, C. M. and Zhuravlev, M. Ye. and Felker, D. and Rzchowski, M. and Eom, C.-B. and Tsymbal, E. Y.},
	year = {2009},
	pages = {3539--3543},
}

@article{boscke_ferroelectricity_2011,
	title = {Ferroelectricity in hafnium oxide thin films},
	volume = {99},
	journal = {Appl. Phys. Lett.},
	author = {Böscke, T. S. and Müller, J. and Bräuhaus, D. and Schröder, U. and Böttger, U.},
	year = {2011},
	pages = {102903},
}

@article{chanthbouala_ferroelectric_2012,
	title = {A ferroelectric memristor},
	volume = {11},
	journal = {Nat. Mater.},
	author = {Chanthbouala, André and Garcia, Vincent and Cherifi, Ryan O. and Bouzehouane, Karim and Fusil, Stéphane and Moya, Xavier and Xavier, Stéphane and Yamada, Hiroyuki and Deranlot, Cyrile and Mathur, Neil D. and Bibes, Manuel and Barthélémy, Agnès and Grollier, Julie},
	year = {2012},
	pages = {860--864},
}

@article{wei_spontaneous_2019,
	title = {Spontaneous polarization enhancement in ferroelectric {$\mathrm{Hf}_{0.5}\mathrm{Zr}_{0.5}\mathrm{O}_{2}$} using atomic oxygen defects engineering: an ab initio study},
	volume = {115},
	journal = {Appl. Phys. Lett.},
	author = {Wei, Wei and Ma, Xiaolei and Wu, Jixuan and Wang, Fei and Zhan, Xuepeng and Li, Yuan and Chen, Jiezhi},
	year = {2019},
	pages = {092905},
}

@article{garcia_ferroelectric_2014,
	title = {Ferroelectric tunnel junctions for information storage and processing},
	volume = {5},
	journal = {Nat. Commun.},
	author = {Garcia, Vincent and Bibes, Manuel},
	year = {2014},
	pages = {4289},
}

@article{park_study_2015,
	title = {Study on the internal field and conduction mechanism of atomic layer deposited ferroelectric {$\mathrm{Hf}_{0.5}\mathrm{Zr}_{0.5}\mathrm{O}_{2}$} thin films},
	volume = {3},
	journal = {J. Mater. Chem. C},
	author = {Park, M. H. and Kim, H. J. and Kim, Y. J. and Moon, T. and Kim, K. D. and Lee, Y. H. and Hyun, S. D. and Hwang, C. S.},
	year = {2015},
	pages = {6291--6300},
}

@article{velev_predictive_2016,
	title = {Predictive modelling of ferroelectric tunnel junctions},
	volume = {2},
	journal = {npj Comput. Mater.},
	author = {Velev, Julian P. and Burton, John D. and Zhuravlev, Mikhail Ye and Tsymbal, Evgeny Y.},
	year = {2016},
	pages = {1--13},
}

@article{boyn_learning_2017,
	title = {Learning through ferroelectric domain dynamics in solid-state synapses},
	volume = {8},
	journal = {Nat. Commun.},
	author = {Boyn, Sören and Grollier, Julie and Lecerf, Gwendal and Xu, Bin and Locatelli, Nicolas and Fusil, Stéphane and Girod, Stéphanie and Carrétéro, Cécile and Garcia, Karin and Xavier, Stéphane and Tomas, Jean and Bellaiche, Laurent and Bibes, Manuel and Barthélémy, Agnès and Saïghi, Sylvain and Garcia, Vincent},
	year = {2017},
	pages = {14736},
}

@article{ambriz-vargas_complementary_2017,
  author  = {Ambriz-Vargas, F. and Kolhatkar, G. and Broyer, M. and Hadj-Youssef, A. and Nouar, R. and Sarkissian, A. and Thomas, R. and Gomez-Yáñez, C. and Gauthier, M. A. and Ruediger, A.},
  title   = {A complementary metal oxide semiconductor process-compatible ferroelectric tunnel junction},
  journal = {ACS Appl. Mater. Interfaces},
  year    = {2017},
  volume  = {9},
  pages   = {13262--13268},
}

@article{theis_end_2017,
	title = {The end of {Moore}'s law: a new beginning for information technology},
	volume = {19},
	journal = {Comput. Sci. Eng},
	author = {Theis, Thomas N. and Wong, H.-S. Philip},
	year = {2017},
	pages = {41--50},
}

@article{perevalov_electronic_2018,
	title = {Electronic structure of stoichiometric and oxygen-deficient ferroelectric {$\mathrm{Hf}_{0.5}\mathrm{Zr}_{0.5}\mathrm{O}_{2}$}},
	volume = {29},
	journal = {Nanotechnology},
	author = {Perevalov, T V and Islamov, D R and Gritsenko, V A and Prosvirin, I P},
	year = {2018},
	pages = {194001},
}

@article{yoon_understanding_2019,
	title = {Understanding tunneling electroresistance effect through potential profile in {$\mathrm{Pt}/\mathrm{Hf}_{0.5}\mathrm{Zr}_{0.5}\mathrm{O}_{2}/\mathrm{TiN}$} ferroelectric tunnel junction memory},
	volume = {115},
	journal = {Appl. Phys. Lett.},
	author = {Yoon, Jungkyu and Hong, Seunghyeon and Song, Yong Won and Ahn, Ji-Hoon and Ahn, Seung-Eon},
	month = oct,
	year = {2019},
	pages = {153502},
}

@article{cheema_enhanced_2020,
	title = {Enhanced ferroelectricity in ultrathin films grown directly on silicon},
	volume = {580},
	journal = {Nature},
	author = {Cheema, Suraj S. and Kwon, Daewoong and Shanker, Nirmaan and dos Reis, Roberto and Hsu, Shang-Lin and Xiao, Jun and Zhang, Haigang and Wagner, Ryan and Datar, Adhiraj and McCarter, Margaret R. and Serrao, Claudy R. and Yadav, Ajay K. and Karbasian, Golnaz and Hsu, Cheng-Hsiang and Tan, Ava J. and Wang, Li-Chen and Thakare, Vishal and Zhang, Xiang and Mehta, Apurva and Karapetrova, Evguenia and Chopdekar, Rajesh V and Shafer, Padraic and Arenholz, Elke and Hu, Chenming and Proksch, Roger and Ramesh, Ramamoorthy and Ciston, Jim and Salahuddin, Sayeef},
	year = {2020},
	pages = {478--482},
}

@article{sebastian_memory_2020,
	title = {Memory devices and applications for in-memory computing},
	volume = {15},
	journal = {Nat. Nanotechnol.},
	author = {Sebastian, Abu and Le Gallo, Manuel and Khaddam-Aljameh, Riduan and Eleftheriou, Evangelos},
	year = {2020},
	pages = {529--544},
}

@article{materano_influence_2020,
	title = {Influence of oxygen content on the structure and reliability of ferroelectric {$\mathrm{Hf}_{x}\mathrm{Zr}_{1-x}\mathrm{O}_{2}$} layers},
	volume = {2},
	journal = {ACS Appl. Electron. Mater.},
	author = {Materano, Monica and Mittmann, Terence and Lomenzo, Patrick D. and Zhou, Chuanzhen and Jones, Jacob L. and Falkowski, Max and Kersch, Alfred and Mikolajick, Thomas and Schroeder, Uwe},
	year = {2020},
	pages = {3618--3626},
}

@INPROCEEDINGS{Mittmann_impact_2020,
  author={Mittmann, T. and Materano, M. and Chang, S.-C. and Karpov, I. and Mikolajick, T. and Schroeder, U.},
  booktitle={{IEDM}}, 
  title={Impact of oxygen vacancy content in ferroelectric {HZO} films on the device performance}, 
  year={2020},
  pages={18.4.1--18.4.4},
  }

@inproceedings{chu_ultra-thin_2021,
	title = {Ultra-thin {$\mathrm{Hf}_{0.5}\mathrm{Zr}_{0.5}\mathrm{O}_{2}$} ferroelectric tunnel junction with high current density},
	booktitle = {{VLSI-TSA} 2021},
	author = {Chu, Yueh-Hua and Huang, Hsin-Hui and Chen, Yu-Hao and Hsu, Chien-Hua and Tzeng, Pei-Jer and Sheu, Shyh-Shyuan and Lo, Wei-Chung and Wu, Chih-I and Hou, Tuo-Hung},
	year = {2021},
	pages = {1--2},
}

@INPROCEEDINGS{das_beyond_2021,
  author={Das, Shamik and Chen, An and Marinella, Matt},
  booktitle={2021 {IEEE} International Roadmap for Devices and Systems Outbriefs}, 
  title={Beyond {CMOS}}, 
  year={2021},
  pages={1--129},
}

@article{chai_first-principles_2022,
	title = {First-principles study of oxygen vacancy defects in orthorhombic {$\mathrm{Hf}_{0.5}\mathrm{Zr}_{0.5}\mathrm{O}_{2}/\mathrm{SiO}_{2}/\mathrm{Si}$} gate stack},
	volume = {132},
	journal = {J. Appl. Phys.},
	author = {Chai, Junshuai and Xu, Hao and Xiang, Jinjuan and Zhang, Yuanyuan and Zhou, Lixing and Zhao, Shujing and Tian, Fengbin and Duan, Jiahui and Han, Kai and Wang, Xiaolei and Luo, Jun and Wang, Wenwu and Ye, Tianchun and Guo, Yuzheng},
	year = {2022},
	pages = {105301},
}

@article{cheema_one_2022,
	title = {One nanometer {$\mathrm{HfO}_{2}$}-based ferroelectric tunnel junctions on silicon},
	volume = {8},
	journal = {Adv. Electron. Mater.},
	author = {Cheema, Suraj S. and Shanker, Nirmaan and Hsu, Cheng-Hsiang and Datar, Adhiraj and Bae, Jongho and Kwon, Daewoong and Salahuddin, Sayeef},
	year = {2022},
	pages = {2100499},
}

@article{fontanini_charge_2022,
  title = {Charge-Trapping-Induced Compensation of the Ferroelectric Polarization in FTJs: Optimal Conditions for a Synaptic Device Operation},
  author = {Fontanini, Riccardo and Segatto, Mattia and Nair, Keerthana S. and Holzer, Marco and Driussi, Francesco and H{\"a}usler, Ingrid and Koch, Christoph T. and Dubourdieu, Catherine and Deshpande, Vikas and Esseni, David},
  journal = {IEEE Trans. Electron Devices},
  volume = {69},
  pages = {3694--3699},
  year = {2022},
}

@article{yao_ferroelectric_2023,
	title = {Ferroelectric {Schottky} diodes of {$\mathrm{CuInP}_{2}\mathrm{S}_{6}$} nanosheet},
	volume = {123},
	journal = {Appl. Phys. Lett.},
	author = {Yao, Jinyuan and Liu, Yongtao and Ding, Shaoqing and Zhu, Yanglin and Mao, Zhiqiang and Kalinin, Sergei V. and Liu, Ying},
	year = {2023},
	pages = {142903},
}

@article{mallick_oxygen_2023,
	title = {Oxygen vacancy contributions to the electrical stress response and endurance of ferroelectric hafnium zirconium oxide thin films},
	volume = {122},
	journal = {Appl. Phys. Lett.},
	author = {Mallick, Antik and Lenox, Megan K. and Beechem, Thomas E. and Ihlefeld, Jon F. and Shukla, Nikhil},
	year = {2023},
	pages = {132902},
}

@article{yoon_review_2023,
	title = {A review on device requirements of resistive random access memory ({RRAM})-based neuromorphic computing},
	volume = {11},
	journal = {APL Mater.},
	author = {Yoon, Jeong Hyun and Song, Young-Woong and Ham, Wooho and Park, Jeong-Min and Kwon, Jang-Yeon},
	year = {2023},
	pages = {090701},
}

@inproceedings{jung_superior_2023,
  author    = {Jung, L. and Lee, J. and Oh, S. and Hwang, H.},
  title     = {Superior Retention ({$>$}1 Year, 85~{$^\circ$C}) and Memory Window ({$\sim$}1.8 V) Using Ultra-Thin {HZO FTJ} with {OTS} Selector for {X}-Point Memory Applications},
  booktitle = {IEDM 2023},
  year      = {2023},
  pages     = {1--4},
}

@article{park_ferroelectric_2024,
	title = {Ferroelectric tunnel junctions: promise, achievements and challenges},
	volume = {57},
	journal = {J. Phys. D: Appl. Phys.},
	author = {Park, Sung Hyuk and Lee, Hyeon Ji and Park, Min Hyuk and Kim, Jihyun and Jang, Ho Won},
	year = {2024},
	pages = {253002},
}

@article{hwang_physics_2024,
	title = {Physics, structures, and applications of fluorite-structured ferroelectric tunnel junctions},
	volume = {20},
	journal = {Small},
	author = {Hwang, Junghyeon and Goh, Youngin and Jeon, Sanghun},
	year = {2024},
	pages = {2305271},
}

@article{beechem_photoinduced_2024,
	title = {Photoinduced patterning of oxygen vacancies to promote the ferroelectric phase of {$\mathrm{Hf}_{0.5}\mathrm{Zr}_{0.5}\mathrm{O}_{2}$}},
	volume = {124},
	journal = {Appl. Phys. Lett.},
	author = {Beechem, Thomas E. and Vega, Fernando and Jaszewski, Samantha T. and Aronson, Benjamin L. and Kelley, Kyle P. and Ihlefeld, Jon F.},
	year = {2024},
	pages = {062902},
}

@article{luo_correlated_2024,
  author  = {Luo, Yuan-Chun and Khanna, Abhishek and Grisafe, Benjamin and Sun, Jiaxin and Dutta, Sourav and Noskin, Lindsey E. and Adamo, Carolina and Mei, Antonio B. and Ghosh, Ram Krishna and Colletta, Michael and Holtz, M. E. and Gambin, Vincent and Kourkoutis, Lena F. and Yu, Shimeng and Schlom, Darrell G. and Datta, Suman},
  title   = {Correlated Oxide Selector for Cross-Point Embedded Non-Volatile Memory},
  journal = {IEEE Trans. Electron Devices},
  volume  = {71},
  pages   = {916--921},
  year    = {2024},
}

@article{vega_modifications_2025,
	title = {Modifications in the charge trap landscape in {$\mathrm{Hf}_{0.5}\mathrm{Zr}_{0.5}\mathrm{O}_{2}$} as a function of oxygen vacancy concentration observed with photoemission electron microscopy},
	volume = {126},
	journal = {Appl. Phys. Lett.},
	author = {Vega, Fernando and Boehm, Alex and Kim, Andrew R. and Jaszewski, Samantha and Ohta, Taisuke and Ihlefeld, Jon and Beechem, Thomas},
	year = {2025},
	pages = {152903},
}

@article{engl_peak_2025,
	title = {Peak splitting and bias fields in ferroelectric hafnia mediated by interface charge effects},
	volume = {17},
	journal = {ACS Appl. Mater. Interfaces},
	author = {Engl, Moritz and Hamouda, Wassim and Häusler, Ines and Lancaster, Suzanne and Carpentieri, Luca and Mikolajick, Thomas and Dubourdieu, Catherine and Slesazeck, Stefan},
	year = {2025},
	pages = {51468--51475},
}

@article{yang_mathrmdftuv_2025,
	title = {{$\mathrm{DFT}+U+V$} study of pristine and oxygen-deficient {$\mathrm{HfO}_{2}$} with self-consistent {Hubbard} parameters},
	volume = {9},
	journal = {Phys. Rev. Mater.},
	author = {Yang, Yudi and Yang, Wooil and Son, Young-Woo and Liu, Shi},
	year = {2025},
	pages = {034402},
}

@article{xiong_silicon-based_2026,
	title = {Silicon-based {Josephson} junction field-effect transistors enabling cryogenic logic and quantum technologies},
	volume = {1},
	journal = {APL Eng. Phys.},
	author = {Xiong, Yusheng and Delfanazari, Kaveh},
	year = {2026},
	pages = {011503},
}

@article{noor_oxygen_2026,
	title = {Oxygen vacancy energetics and kinetics at {$\mathrm{Hf}_{0.5}\mathrm{Zr}_{0.5}\mathrm{O}_{2}$}--{$\mathrm{WO}_{3}$} interfaces: impact on ferroelectric behavior and device functionality},
	volume = {8},
	journal = {ACS Appl. Electron. Mater.},
	author = {Noor, Manifa and Kim, Jongchan and Bergschneider, Matthew and Kummel, Andrew C. and Cho, Kyeongjae},
	year = {2026},
	pages = {824--833},
}

@article{ding_indirect_2026,
	title = {Indirect tunneling enabled spontaneous time-reversal symmetry breaking and {Josephson} diode effect in {$\mathrm{TiN}/\mathrm{Al}_{2}\mathrm{O}_{3}/\mathrm{Hf}_{0.8}\mathrm{Zr}_{0.2}\mathrm{O}_{2}/\mathrm{Nb}$} tunnel junctions},
	volume = {113},
	journal = {Phys. Rev. B},
	author = {Ding, Shaoqing and Yao, Jinyuan and Bi, Zhen and Tran, Quyen and Liu, Bangzhi and Li, Qi and Trolier-McKinstry, Susan and Jackson, Thomas N. and Liu, Ying},
	year = {2026},
	pages = {134510},
}

@book{Neamen_SM_2012,
  title={Semiconductor Physics and Devices: Basic Principles},
  author={Neamen, Donald A.},
  year={2012},
  pages = {631},
  publisher={McGraw-Hill, \& The McGraw-Hill Companies, Inc.}
}

@article{Altendorf_metallicWO3_2016,
    author = {Altendorf, Simone G. and Jeong, Jaewoo and Passarello, Donata and Aetukuri, Nagaphani B. and Samant, Mahesh G. and Parkin, S. S.},
    title = {Facet-Independent Electric-Field-Induced Volume Metallization of Tungsten Trioxide Films},
    journal = {Advanced Materials},
    year = {2016},
    volume = {28},
    page = {5284-5292}
}

@article{Leng_metallicWO3_2017,
    author = {Leng, Xiang and Pereiro, Juan and Strle, Juan and Dubuis, G. and Bollinger, A. T. and Gozar, A. and Wu, Wu, J. and Litombe, N. and Panagopoulos, C. and Pavuna, D. and Božović, I.},
    title = {Insulator to metal transition in WO3 induced by electrolyte gating},
    journal = {npj Quantum Materials},
    year = {2017},
    volume = {2},
    page = {35}
}

\end{document}